\documentclass[pra,aps,twocolumn,superscriptaddress,showkeys,nofootinbib]{revtex4-1}
\usepackage[latin1]{inputenc}
\usepackage{amsmath,amsthm,amssymb,graphicx,subfigure,xcolor}
\usepackage{mathrsfs}
\usepackage{graphics,float}
\usepackage{epstopdf}
\usepackage{color}
\usepackage[unicode=true]{hyperref}
\usepackage{booktabs}
\usepackage{tabularx}
\usepackage{tabu}
\usepackage{multirow}
\allowdisplaybreaks[4]
\hypersetup{
     colorlinks=true,       		
     linkcolor=blue,          	
     citecolor=blue,            
     urlcolor=blue,           	
 }

\newtheorem*{theorem*}{Theorem}

\newtheorem*{corollary*}{Corollary}

\newtheorem*{lemma*}{Lemma}

\newtheorem*{proposition*}{Proposition}
\theoremstyle{definition}

\newtheorem*{definition*}{Definition}
\theoremstyle{remark}

\newtheorem*{remark*}{Remark}

\begin{document}

\title{Parameterized multipartite entanglement measures}

\author{Hui Li}
\author{Ting Gao}
\email{gaoting@hebtu.edu.cn}
\affiliation{School of Mathematical Sciences, Hebei Normal University, Shijiazhuang 050024, China}
\affiliation{Hebei Mathematics Research Center,  Hebei Normal University, Shijiazhuang 050024, China}
\affiliation{Hebei International Joint Research Center for Mathematics and Interdisciplinary Science, Hebei Normal University, Shijiazhuang 050024, China}
\author{Fengli Yan}
\email{flyan@hebtu.edu.cn}
\affiliation{College of Physics, Hebei Key Laboratory of Photophysics Research and Application, Hebei Normal University, Shijiazhuang 050024, China}
\begin{abstract}
We investigate parameterized multipartite entanglement measures from the perspective of $k$-nonseparability in this paper. We present two types of entanglement measures in $n$-partite systems, $q$-$k$-ME concurrence $(q\geq2,~2\leq k\leq n)$ and $\alpha$-$k$-ME concurrence $(0\leq\alpha\leq\frac{1}{2},~2\leq k\leq n)$, which unambiguously detect all $k$-nonseparable states in arbitrary $n$-partite systems. Rigorous proofs show that the proposed $k$-nonseparable measures satisfy all the requirements for being an entanglement measure including the entanglement monotone, strong monotone, convexity, vanishing on all $k$-separable states, and being strictly greater than zero for all $k$-nonseparable states. In particular, the $q$-2-ME concurrence and $\alpha$-2-ME concurrence, renamed as $q$-GME concurrence and $\alpha$-GME concurrence, respectively, are two kinds of genuine entanglement measures corresponding the case where the systems are divided into bipartition $(k=2)$. The lower bounds of two classes $k$-nonseparable measures are obtained by employing the approach that takes into account the permutationally invariant part of a quantum state. And the relations between $q$-$n$-ME concurrence ($\alpha$-$n$-ME concurrence) and global negativity are established. In addition, we discuss the degree of separability and elaborate on an effective detection method with concrete examples. Moreover, we compare the $q$-GME concurrence defined by us to other genuine entanglement measures.
~\\
\end{abstract}



\maketitle

\section{Introduction}
Quantum entanglement as a physical resource is indispensable in such tasks as  quantum cryptography \cite{7,8,9,10}, quantum teleportation \cite{11,12,13}, and quantum communication \cite{14,15,6}. Moreover, it is recognized that entangled states  are at the core of quantum information processing.  Therefore, the qualitative and quantitative study of multipartite quantum states is a matter of great importance, and in this paper we mainly focus on the quantitative description of entanglement of states.

Initially, bipartite systems have been studied extensively, and a wide range of measures have been found to quantify the entanglement of states. Concurrence is one of well-known measures for bipartite quantum systems \cite{22,23,24,25}, and Wootters gave an analytical expression for arbitrary 2-qubit quantum states in Ref. \cite{24}. Furthermore, there are other methods, such as negativity \cite{26,27}, entanglement of formation \cite{28,29}, Tsallis entropy of entanglement \cite{30}, that can also characterize the entanglement of quantum states commendably.

Many efforts have been made to detect multipartite entanglement \cite{3,31,32,33,34,35,36,17,18,19,39}, but no measure can be employed to calculate the entanglement of multipartite mixed states. In Ref. \cite{20}, Ma $et~al$. put forth a measure of genuine multipartite entanglement (GME), termed GME concurrence, which can distinguish the genuinely entangled states from the others. Since computing the entanglement of a state involves optimization procedures that are considerably more difficult to handle, they rendered a computable lower bound. Subsequently, Chen $et~al.$ \cite{21} optimized the lower bound of Ref. \cite{20}. In order to  quantitatively characterize the whole hierarchy of $k$-separability of states more precisely in $n$-partite systems, Hong $et~al$. \cite{1} advanced generalized measures called $k$-ME concurrence, where $k$ runs from $n$ to 2, and provided their two strong lower bounds. The GME concurrence \cite{20} is a special case of the $k$-ME concurrence \cite{1} when $k=2$. It is acknowledged that multipartite entanglement (ME) is extremely complicated, Gao $et~al$. \cite{2} proposed that whether a state is $k$-nonseparable can be determined by its permutationally invariant (PI) part, which dramatically reduces the dimension of the space to be considered.

More recently, there are several works on the genuine entanglement measures. In 2021, Xie and Eberly \cite{42} put forward a genuine entanglement measure for 3-qubit quantum systems, concurrence fill, which is obtained based on Heron formula of triangle area. Later, Li and Shang \cite{41} also advanced a new measure called geometric mean of bipartite concurrences (GBC), which does not involve the minimization procedure in calculating the entanglement of pure states.

In addition, some researchers devoted themselves to the study of parameterized measures. Yang $et~al$. \cite{5} introduced a parameterized entanglement monotone \cite{4} called $q$-concurrence $(q\geq2)$ for arbitrary bipartite systems, and presented the lower bound of $q$-concurrence meanwhile. Later, Wei and Fei \cite{16} came up with a generalized concurrence in terms of different ranges of parameter named $\alpha$-concurrence $(0\leq\alpha\leq\frac{1}{2})$. Shi \cite{40} generalized the measure GBC defined by Li and Shang \cite{41} to parameterized concurrence, which is known as geometric mean of $q$-concurrence (G$q$C).

Motivated by the thoughts in Refs. \cite{1,2,5,16}, our aim in this paper is to define new ME measures from the point of $k$-nonseparability utilizing the parameterized concurrence. The content of this paper is arranged as follows. In Sec. \ref{II}, we briefly introduce a few notions. In Sec. \ref{III}, we put forward two classes of parameterized measures called $q$-$k$-ME concurrence $C_{q-k}$ $(q\geq2)$ and $\alpha$-$k$-ME concurrence $C_{\alpha-k}$ $(0\leq\alpha\leq\frac{1}{2})$, respectively. When $k=2$, the corresponding two categories of genuine entanglement measures, respectively, are termed $q$-GME concurrence $(C_{q-\rm GME})$ and $\alpha$-GME concurrence $(C_{\alpha-\rm GME})$. We verify that these measures defined by us conform with, simultaneously, the properties including non-negativity, being strictly greater than zero for all $k$-nonseparable states, invariance under local unitary transformations, (strong) monotonicity, and convexity. In addition, $C_{q-k}$ and $C_{q-\rm GME}$ satisfy the subadditivity as well, but $C_{\alpha-k}$ and $C_{\alpha-\rm GME}$ fail. The lower bound of $C_{q-k}$ is obtained in Sec. \ref{IV} by taking the maximum of $C_{q-k}$ of the PI part of a quantum state, and so is $C_{\alpha-k}$. Furthermore, we establish the relations between $C_{q-n}$ ($C_{\alpha-n}$) and global negativity, which can be used to detect whether a quantum state is entangled or fully separable. Meanwhile, we discuss a specific example of mixing W state with white noise, and observe that the range detected by our results is larger than that of Ref. \cite{19}. These two approaches, reflecting the degree of entanglement in Sec. \ref{V}, exhibit different aspects of dominance in the detection of entanglement. The combination of these two methods can be used to detect $k$-nonseparable states more effectively.  In Sec. \ref{VI}, we compare $C_{q-\rm GME}$ with concurrence fill \cite{42} and G$q$C \cite{40} by a concrete example, which shows that they generate different entanglement orders as well as $C_{q-\rm GME}$ is smoother at times. We conclude in Sec. \ref{VII}.

\section{Preliminaries}\label{II}
Let $\rho$ be an $n$-partite quantum state on Hilbert space $\mathcal{H}=\mathcal{H}_{1}\otimes\mathcal{H}_{2}\otimes \cdots\otimes\mathcal{H}_{n}$ with dim${\mathcal{H}_{i}}=d_i$, and $A_1|A_2|\cdots|A_k$ be a $k$-partition $(2\leq k\leq n)$ of set $A=\{1,2,\cdots,n\}$ such that
\begin{equation}\label{1}
\begin{array}{rl}
\bigcup_{t=1}^{k}A_{t}=\{1,2,\cdots,n\},A_{t}\bigcap A_{t'}=\emptyset~{\rm when}~t\neq t'.
\end{array}
\end{equation}

An $n$-partite pure state $|\varphi\rangle$ on Hilbert space $\mathcal{H}$ is referred to as $k$-separable ($2\leq k\leq n$) \cite{25} if there exists a splitting of the $n$ parties into $k$ parts $A_1,A_2,\cdots,A_k$ such that $|\varphi\rangle=\otimes_{t=1}^{k}|\varphi_t\rangle_{A_t}$ holds. An $n$-partite mixed state $\rho$ is called $k$-separable if it can be represented as a convex combination of $k$-separable pure states, that is, $\rho=\sum_{i}p_i|\varphi_i\rangle\langle\varphi_i|$, where  $|\varphi_i\rangle$ could be $k$-separable regarding different partitions fulfilling the above condition (\ref{1}). Otherwise, the quantum state is called $k$-nonseparable. If $\rho$ is $n$-separable, then it is called fully separable. If not, it is said to be entangled.

For any $n$-partite quantum state $\rho$, its PI part can be denoted as \cite{2}
\begin{equation}\label{2}
\begin{array}{rl}
\rho^{\rm PI}=\frac{1}{n!}\sum_{j=1}^{n!}\Pi_j\rho\Pi_j^\dagger,
\end{array}
\end{equation}
where the set $\{\Pi_j\}$ contains all of permutations of $n$ particles.

A well-defined $k$-nonseparability measure $E(\rho)$ ought to meet the conditions as follows:

(M1) $E(\rho)=0$ for arbitrary $k$-separable quantum states.

(M2) $E(\rho)>0$ for arbitrary $k$-nonseparable quantum states.

(M3) (Invariance under local unitary transformations) $E(\rho)=E(U_{\rm Local}\rho U_{\rm Local}^{\dagger})$.

(M4) (Monotonicity) $E$ is non-increasing under local operations and classical communication (LOCC), i.e., $E(\Lambda_{\rm LOCC}(\rho))\leq E(\rho)$. Moreover, several entanglement measures can obey a stronger condition called strong monotonicity, that $E$ is average non-increasing under LOCC, i.e., $E(\rho)\geq\sum_jp_jE(\sigma_j)$ with $\{p_j,\sigma_j\}$ being yielded after $\Lambda_{\rm LOCC}$ acts on state $\rho$.

(M5) Most entanglement measures also conform to convexity, $E(\sum_{i}p_{i}\rho_{i})\leq \sum_{i}p_{i}E(\rho_{i})$.

(M6) Further there may be several entanglement measures that satisfy subadditivity, $E(\rho\otimes\sigma)\leq E(\rho)+E(\sigma)$.

For any bipartite pure state $|\varphi\rangle_{AB}$, Yang $et~al$. \cite{5} defined a parameterized bipartite entanglement measure, $q$-concurrence $(q\geq2)$, which is
\begin{equation}\label{3}
\begin{array}{rl}
C_{q}(|\varphi\rangle_{AB})=1-{\rm Tr}(\rho_A^q).
\end{array}
\end{equation}
And then, Wei and Fei \cite{16} also came up with a new bipartite entanglement measure in terms of different parameter ranges, $\alpha$-concurrence $(0\leq\alpha\leq\frac{1}{2})$, the form is
\begin{equation}\label{4}
\begin{array}{rl}
C_{\alpha}(|\varphi\rangle_{AB})={\rm Tr}(\rho_A^\alpha)-1.
\end{array}
\end{equation}
Here $\rho_A$ is the reduced density operator of $|\varphi\rangle_{AB}$.

Our purpose is to generalize them to multipartite quantum systems in the following sections.

\section{New parameterized entanglement measures}\label{III}
We start by introducing new measures, $q$-$k$-ME concurrence ($q\geq2$) and $\alpha$-$k$-ME concurrence $(0\leq\alpha\leq\frac{1}{2})$, which are inspired by those proposed in Refs. \cite{1,2,5,16}.

\textbf{Definition 1}. For any $n$-partite pure state $|\varphi\rangle\in \mathcal{H}$, we define the $q$-$k$-ME concurrence as
\begin{equation}\label{5}
\begin{array}{rl}
C_{q-k}(|\varphi\rangle)=\min\frac{\sum\limits_{t=1}^{k}[1-{\rm Tr}(\rho_{A_t}^q)]}{k}
\end{array}
\end{equation}
for any $q\geq2$, and the $\alpha$-$k$-ME concurrence as
\begin{equation}\label{6}
\begin{array}{rl}
C_{\alpha-k}(|\varphi\rangle)=\min\frac{\sum\limits_{t=1}^{k}[{\rm Tr}(\rho_{A_t}^\alpha)-1]}{k}
\end{array}
\end{equation}
for any $0\leq\alpha\leq\frac{1}{2}$, respectively. Here $\rho_{A_t}$ is the reduced density operator of subsystem $A_t$, and the minimum is done over all feasible $k$-partitions obeying condition (\ref{1}).

For any $n$-partite mixed state $\rho$, the $q$-$k$-ME concurrence is defined as
\begin{equation}\label{07}
\begin{array}{rl}
C_{q-k}(\rho)=\inf\limits_{\{p_i,|\varphi_{i}\rangle\}}\sum\limits_{i}p_iC_{q-k}(|\varphi_i\rangle)
\end{array}
\end{equation}
for any $q\geq2$. Analogously, the $\alpha$-$k$-ME concurrence is defined as
\begin{equation}\label{08}
\begin{array}{rl}
C_{\alpha-k}(\rho)=\inf\limits_{\{p_i,|\varphi_{i}\rangle\}}\sum\limits_{i}p_iC_{\alpha-k}(|\varphi_i\rangle)
\end{array}
\end{equation}
for any $0\leq\alpha\leq\frac{1}{2}$. Here the infimum is taken over all viable pure decompositions $\rho=\sum_{i}p_i|\varphi_i\rangle\langle\varphi_i|$.

It is worth noting that ${\rm GME}$ concurrence $C_{\rm GME}$ \cite{20}, $k$-ME concurrence $C_{k-\rm ME}$ \cite{1}, and $q$-concurrence $C_q$ \cite{5} are the special cases of $q$-$k$-ME concurrence in some sense. For any pure state $|\varphi\rangle$, when $q=2$, $C_{2-k}(|\varphi\rangle)=\frac{C_{k-{\rm ME}}^2(|\varphi\rangle)}{2}$; when $q=k=2$, $C_{2-2}(|\varphi\rangle)=\frac{C_{\rm GME}^2(|\varphi\rangle)}{2}$; when $n=2$, $C_{q-2}(\rho)$ and $C_q(\rho)$ are equal for any quantum state $\rho$. The bipartite entanglement measure, $\alpha$-concurrence $C_{\alpha}$ \cite{16}, is the special case of $\alpha$-$k$-ME concurrence. $C_{\alpha-2}(\rho)$ is consistent with $C_{\alpha}(\rho)$ when $n=2$.

The following we will verify $q$-$k$-ME concurrence ($q\geq2$) and $\alpha$-$k$-ME concurrence ($0\leq\alpha\leq\frac{1}{2}$) satisfy the requirements of an entanglement measure.

\textbf{Proposition 1}. The $q$-$k$-ME concurrence ($q\geq2$) and $\alpha$-$k$-ME concurrence ($0\leq\alpha\leq\frac{1}{2}$) are well-defined entanglement measures.

\textbf{Proof}. First, we prove $q$-$k$-ME concurrence ($q\geq2$) satisfies the conditions (M1)---(M6).

(M1) Since ${\rm Tr}\rho^q\leq1$, we can know $C_{q-k}(\rho)\geq0$. Given a $k$-separable pure state $|\varphi\rangle=\otimes_{t=1}^{k}|\varphi_t\rangle_{A_t}$, $A_{1}$ and $\overline{A}_{1}$, $A_{2}$ and $\overline{A}_{2}$, $\cdots,$ $A_{k}$ and $\overline{A}_{k}$  don't exist correlation, where $\overline{A}_{i}$ is the complement of $A_{i}$, one obtains $C_{q}(|\varphi\rangle_{A_t|\overline{A}_t})=0$, $t=1,2,\cdots,k$. Then we can easily get $C_{q-k}(|\varphi\rangle)=0$ for arbitrary $k$-separable pure states. For any $k$-separable mixed state $\rho$ with pure state ensemble decomposition $\{p_i,\rho_{i}\}$, $\rho_{i}=|\varphi_i\rangle\langle\varphi_i|$ and $|\varphi_i\rangle$ is $k$-separable, then $C_{q-k}(\rho)\leq\sum_{i}p_iC_{q-k}(|\varphi_i\rangle)=0$. Therefore, $C_{q-k}(\rho)=0$ for any $k$-separable quantum state.

(M2) Suppose that the set \{$A_1|A_2|\cdots|A_k$\} contains all of $k$-partitions $(2\leq k\leq n)$ of set $A=\{1,2,\cdots,n\}$. For arbitrary $k$-nonseparable pure states, there exist $k'\in\{1,2,\cdots,k\}$ such that subsystems $A_{k'}$ and $\overline{A}_{k'}$ are entangled, then $C_q(|\varphi\rangle_{A_{k'}|\overline{A}_{k'}})=1-{\rm Tr}(\rho_{A_{k'}}^q)>0$, thus we can easily derive $C_{q-k}(|\varphi\rangle)>0$. For any $k$-nonseparable mixed state $\rho$, there is no convex combination of $k$-separable pure states. Hence $C_{q-k}(\rho)>0$ for any $k$-nonseparable state.

(M3) By the property of trace, $C_{q-k}(\rho)$ is invariant under local unitary transformation.

(M4) We first demonstrate that $C_{q-k}$ satisfies monotonicity.

Because $q$-concurrence is non-increasing under LOCC for arbitrary bipartite pure states \cite{5}, namely, the inequality $C_{qA_t|\overline{A}_t}(\Lambda_{\rm LOCC}(|\varphi\rangle))\leq C_{qA_t|\overline{A}_t}(|\varphi\rangle)$ holds for any pure state $|\varphi\rangle$, then we can get
\begin{equation*}
\begin{array}{rl}
&C_{q-k}(\Lambda_{\rm LOCC}(|\varphi\rangle))\\
=&\min\frac{\sum\limits_{t=1}^{k}C_{q A_t|{\overline A}_t}(\Lambda_{\rm LOCC}(|\varphi\rangle))}{k}\\
\leq&\min\frac{\sum\limits_{t=1}^{k}C_{q A_t|{\overline A}_t}(|\varphi\rangle)}{k}\\
=&C_{q-k}(|\varphi\rangle).\\
\end{array}
\end{equation*}

For any mixed state $\rho$ with the optimal pure decomposition $\{p_i, \rho_{i}\}$, $\rho_{i}=|\varphi_i\rangle\langle\varphi_i|$, one has
\begin{equation*}
\begin{array}{rl}
&C_{q-k}(\Lambda_{\rm LOCC}(\rho))\\
\leq&\sum_{i}p_iC_{q-k}(\Lambda_{\rm LOCC}(|\varphi_i\rangle))\\
\leq&\sum_{i}p_iC_{q-k}(|\varphi_i\rangle)\\
=&C_{q-k}(\rho).
\end{array}
\end{equation*}
Here the first inequality is due to the definition of $C_{q-k}(\rho)$, the second inequality holds because $C_{q-k}$ is non-increasing for any pure state under LOCC.

Next we show $C_{q-k}$ conforms to strong monotonicity.

Owing to the fact that $q$-concurrence $(q\geq2)$ is entanglement monotone for arbitrary bipartite quantum states \cite{5}, that is, the inequality $C_{q A_{t}|{\overline A}_{t}}(\rho)\geq \sum_j p_j C_{qA_{t}|{\overline A}_{t}}(\sigma_{j})$ holds, where an ensemble of state $\sigma_j$ with respectively corresponding probability $p_j$ is  obtained by LOCC acting on $\rho$. We first consider the case that $\rho=|\varphi\rangle\langle\varphi|$ and $\sigma_{j}$ are pure states. Assume that $A_1|A_2|\cdots|A_k$ is the optimal partition of $\rho$, then we get
\begin{equation*}
\begin{array}{rl}
C_{q-k}(\rho)&=\frac{\sum\limits_{t=1}^k(1-{\rm Tr}\rho_{A_t}^q)}{k}\\
&=\frac{\sum\limits_{t=1}^kC_{qA_{t}|{\overline A}_{t}}(|\varphi\rangle)}{k}\\
&\geq\frac{\sum\limits_{t=1}^k\sum\limits_{j}p_jC_{qA_{t}|{\overline A}_{t}}(\sigma_j)}{k}\\
&=\sum_{j}p_j\frac{\sum\limits_{t=1}^kC_{qA_{t}|{\overline A}_{t}}(\sigma_j)}{k}\\
&\geq\sum_{j}p_jC_{q-k}(\sigma_j),\\
\end{array}
\end{equation*}
where the last inequality holds according to Eq. (\ref{5}).

For any mixed state $\rho$ with the optimal pure decomposition $\{p_i, \rho_{i}\}$, $\rho_{i}=|\varphi_i\rangle\langle\varphi_i|$, one has
\begin{equation*}
\begin{array}{rl}
C_{q-k}(\rho)&=\sum_ip_iC_{q-k}(|\varphi_{i}\rangle)\\
&\geq\sum_{ij}p_ip(j|i)C_{q-k}(|\varphi_{i}^j\rangle)\\
&=\sum_{ij}p_jp(i|j)C_{q-k}(|\varphi_{i}^j\rangle)\\
&=\sum_{j}p_j(\sum_ip(i|j)C_{q-k}(|\varphi_{i}^j\rangle))\\
&\geq\sum_jp_jC_{q-k}(\sigma_j).\\
\end{array}
\end{equation*}
Here the state $|\varphi_{i}^j\rangle=\frac{\Lambda_j|\varphi_i\rangle}{{\sqrt {{\rm Tr}(\Lambda_j|\varphi_i\rangle\langle\varphi_i|\Lambda_j^\dag)}}}$ is obtained with probability $p(j|i)={{{\rm Tr}(\Lambda_j|\varphi_i\rangle\langle\varphi_i|\Lambda_j^\dag)}}$ after performing stochastic LOCC on $|\varphi_i\rangle$, and $p_{j}={\rm Tr}(\Lambda_j\rho\Lambda_j^\dag)$ is the probability of occurring the outcome $j$ with $\sigma_j=\sum_ip(i|j)|\varphi_{i}^j\rangle\langle\varphi_{i}^j|$, $p(i|j)={p_ip(j|i)}/{p_j}$. The first inequality is true because $C_{q-k}$ satisfies the strong monotonicity for any pure state, while the second inequality holds following from Eq. (\ref{07}).

(M5) The convexity holds due to convex-roof extension.

Further we prove $C_{q-k}$ fulfills the subadditivity.

(M6) Let $\rho$ and $\sigma$ be two arbitrary pure states, and $\rho=|\varphi\rangle\langle\varphi|$, $\sigma=|\phi\rangle\langle\phi|$. Suppose that there exist the optimal partitions $A_1|A_2|\cdots|A_k$ and $B_1|B_2|\cdots|B_k$ satisfying the condition of $k$-partition such that $C_{q-k}(\rho)=\frac{\sum_{t=1}^{k}[1-{\rm Tr}(\rho_{A_t}^q)]}{k}$, $C_{q-k}(\sigma)=\frac{\sum_{t=1}^{k}[1-{\rm Tr}(\sigma_{B_t}^q)]}{k}$, we can get
\begin{equation}\label{7}
\begin{array}{rl}
&C_{q-k}(\rho\otimes\sigma)-C_{q-k}(\rho)-C_{q-k}(\sigma)\\
\leq &\frac{1}{k}\{\sum_{t=1}^{k}[1-{\rm Tr}(\rho_{A_t}^{q}){\rm Tr}(\sigma_{B_t}^{q})]-\sum_{t=1}^{k}[1-{\rm Tr}(\rho_{A_t}^{q})]-\\
&\sum_{t=1}^{k}[1-{\rm Tr}(\sigma_{B_t}^{q})]\}\\
=&\frac{1}{k}\sum_{t=1}^{k}[-{\rm Tr}(\rho_{A_t}^{q}){\rm Tr}(\sigma_{B_t}^{q})+{\rm Tr}(\rho_{A_t}^{q})+{\rm Tr}(\sigma_{B_t}^{q})-1]\\
=&-\frac{1}{k}\sum_{t=1}^{k}[1-{\rm Tr}(\rho_{A_t}^{q})][1-{\rm Tr}(\sigma_{B_t}^{q})]\\
\leq&0.\\
\end{array}
\end{equation}

Suppose that $\rho$ is any mixed state with optimal pure decomposition $\rho=\sum_{i}p_{i}\rho_{i}$ and $\rho_{i}=|\varphi_i\rangle\langle\varphi_i|$, $\sigma=|\phi\rangle\langle\phi|$ is any pure state, then we have
\begin{equation}\label{8}
\begin{array}{rl}
C_{q-k}(\rho\otimes\sigma)&=C_{q-k}(\sum_{i}p_{i}\rho_{i}\otimes|\phi\rangle\langle\phi|)\\
&\leq\sum_{i}p_{i}C_{q-k}(|\varphi_i\rangle\langle\varphi_i|\otimes|\phi\rangle\langle\phi|)\\
&\leq\sum_{i}p_{i}[C_{q-k}(|\varphi_i\rangle)+C_{q-k}(|\phi\rangle)]\\
&=C_{q-k}(\rho)+C_{q-k}(\sigma),
\end{array}
\end{equation}
where the first inequality is owing to the convexity of $C_{q-k}$, and the second inequality can be gotten from the result of inequality (\ref{7}).

By similar procedures, if $\rho,~\sigma$ are any two mixed states, and they have optimal pure decompositions $\rho=\sum_{i}p_{i}|\varphi_{i}\rangle\langle\varphi_{i}|$, $\sigma=\sum_{j}q_{j}|\phi_{j}\rangle\langle\phi_{j}|$, one has
\begin{equation*}
\begin{array}{rl}
C_{q-k}(\rho\otimes\sigma)&=C_{q-k}(\sum_{i}p_{i}\rho_{i}\otimes\sum_{j}q_{j}\sigma_{j})\\
&\leq\sum_{j}q_{j}C_{q-k}(\sum_{i}p_{i}|\varphi_i\rangle\langle\varphi_i|\otimes|\phi_j\rangle\langle\phi_j|)\\
&\leq\sum_{j}q_{j}[C_{q-k}(\rho)+C_{q-k}(\sigma_j)]\\
&=C_{q-k}(\rho)+C_{q-k}(\sigma).
\end{array}
\end{equation*}
Here the first inequality is due to the convexity of $C_{q-k}$, the second inequality holds because of the inequality (\ref{8}).

With similar methods, we can also prove $\alpha$-$k$-ME concurrence $(0\leq\alpha\leq\frac{1}{2})$ meets these requirements (M1)---(M5) for being a ME measure.  And it is not difficult to prove that $C_{\alpha-k}$ does not satisfy subadditivity. $\hfill\blacksquare$

A quantum state is genuinely multipartite entangled iff it is 2-nonseparable. In particular, for the special case $k=2$, the formula (\ref{5}) can be written as
\begin{equation}\label{14}
\begin{array}{rl}
C_{q-{\rm GME}}(|\varphi\rangle)=\min\limits_{\gamma_i\in\gamma}{[1-{\rm Tr}(\rho_{A_{\gamma_i}}^q)]},
\end{array}
\end{equation}
which is called $q$-GME concurrence for any $q\geq2$. Similarly, the formula (\ref{6}) can be reduced to
\begin{equation}\label{15}
\begin{array}{rl}
C_{\alpha-{\rm GME}}(|\varphi\rangle)=\min\limits_{\gamma_i\in\gamma}{[{\rm Tr}(\rho_{\gamma_i}^\alpha)-1]},
\end{array}
\end{equation}
which is termed $\alpha$-GME concurrence for any $0\leq\alpha\leq\frac{1}{2}$. Here $\gamma=\{\gamma_i\}$ expresses the set of all feasible bipartitions.

For an arbitrary $n$-partite mixed state $\rho$, the $q$-GME concurrence is
\begin{equation}\label{16}
\begin{array}{rl}
C_{q-{\rm GME}}(\rho)=\inf\limits_{\{p_i,|\varphi_{i}\rangle\}}\sum\limits_{i}p_iC_{q-{\rm GME}}(|\varphi_i\rangle)
\end{array}
\end{equation}
for any $q\geq2$. Analogously, the $\alpha$-GME concurrence is
\begin{equation}\label{17}
\begin{array}{rl}
C_{\alpha-{\rm GME}}(\rho)=\inf\limits_{\{p_i,|\varphi_{i}\rangle\}}\sum\limits_{i}p_iC_{\alpha-{\rm GME}}(|\varphi_i\rangle)
\end{array}
\end{equation}
for any $0\leq\alpha\leq\frac{1}{2}$, where the infimum runs over all feasible pure decompositions of $\rho$.

The expressions (\ref{16}) and (\ref{17}) are particular cases of the formulas (\ref{07}) and (\ref{08}), respectively. It is natural that $q$-GME concurrence and  $\alpha$-GME concurrence satisfy the necessary requirements (M1)---(M5). Moreover, $q$-GME concurrence fulfills additivity as well. So they can be used to detect whether a quantum state is genuinely multipartite entangled.

Note that $q$-$k$-ME concurrence $C_{q-k}(\rho)\rightarrow1$ as $q\rightarrow+\infty$ when the quantum state $\rho$ is $k$-nonseparable. However, when $\alpha=0$ and the state $\rho$ is $k$-nonseparable, $0$-$k$-ME concurrence $C_{0-k}$ depends on the rank of reduced density operator $\rho_{A_t}$, $C_{0-k}(\rho)=\frac{\sum_{t=1}^{k}r_{A_t}}{k}-1$, where $r_{A_t}$ denotes the rank of $\rho_{A_t}$, $t=1,2,\cdots,k$. Thus, the two entanglement measures $C_{q-k}$ and $C_{\alpha-k}$ describe different aspects. When $q\rightarrow+\infty$, $C_{q-k}$ will take two extremes, $C_{q-k}\rightarrow1$ for any $k$-nonseparable state, whereas $C_{q-k}=0$ for any $k$-separable state, which mean that the states are classified in terms of whether or not they are $k$-separable, if the state is $k$-nonseparable, the degree of entanglement is unified to $1$, otherwise, the degree of entanglement is 0.

Since $q$-concurrence and Tsallis-$q$ entanglement are equivalent for some particular $q$, the $q$-GME concurrence can be viewed as a sort of generalization of $q$-concurrence \cite{5} and Tsallis-$q$ entanglement \cite{30} in some sense. It may provide a method to estimate Tsallis entanglement with a particular parameter $q$ for any multipartite quantum state.

\section{Lower bounds of $q$-$k$-concurrence and $\alpha$-$k$-concurrence}\label{IV}
Compared with the bipartite systems, the structure of the multipartite systems is rather complicated. As a result, it is extremely difficult to give an analytical lower bound since the optimization procedure is involved in computing entanglement of multipartite quantum states. Therefore, we first employ the approach proposed by Gao $et~al$. in Ref. \cite{2}, considering the PI part of quantum state $\rho$, to give the lower bounds of $q$-$k$-ME concurrence $(q\geq2)$ and $\alpha$-$k$-ME concurrence $(0\leq\alpha\leq\frac{1}{2})$.

\textbf{Theorem 1}. For any $n$-partite quantum state $\rho$, the $q$-$k$-ME concurrence $C_{q-k}(\rho)$ ($q\geq2)$ is lower bounded by the maximum of $q$-$k$-ME concurrence of $\rho_U^{\rm PI}$,
\begin{equation}\label{018}
\begin{array}{rl}
C_{q-k}(\rho)\geq\max\limits_U C_{q-k}(\rho_U^{\rm PI}).
\end{array}
\end{equation}
Analogically, the $\alpha$-$k$-ME concurrence $(0\leq\alpha\leq\frac{1}{2})$ satisfies the relation as follows,
\begin{equation}\label{13}
\begin{array}{rl}
C_{\alpha-k}(\rho)\geq\max\limits_U C_{\alpha-k}(\rho_U^{\rm PI}).
\end{array}
\end{equation}
Here the maximum is taken all locally unitary transformations $U$.

\textbf{Proof}. Suppose that the set $\{1,2,\cdots,n\}$ is divided into $A_1|A_2|\cdots|A_k$ satisfying the condition (\ref{1}), then $\Pi_{j}(A_1)|\Pi_{j}(A_2)|\cdots|\Pi_{j}(A_k)$ is still a $k$-partition of the set $\{1,2,\cdots,n\}$. Let $|\varphi\rangle$ be any pure state, then $\Pi_{j}(|\varphi\rangle)$ is also a pure state, and we have
\begin{equation}\label{20}
\begin{array}{rl}
C_{q-k}(|\varphi\rangle)=C_{q-k}(\Pi_j|\varphi\rangle),
\end{array}
\end{equation}
where $\Pi_{j}\in S_n$, $S_n$ is $n$-order symmetric group.

By using the convexity of $C_{q-k}$ and the relation shown in Eq. (\ref{20}), one gets
\begin{equation}\label{21}
\begin{array}{rl}
C_{q-k}(\rho^{\rm PI})&\leq \frac{1}{n!}\sum\limits_{j=1}^{n!}C_{q-k}(\Pi_j|\varphi\rangle)\\
&=\frac{1}{n!}\sum\limits_{j=1}^{n!}C_{q-k}(|\varphi\rangle)=C_{q-k}(|\varphi\rangle).
\end{array}
\end{equation}

Given a mixed state $\rho$, assume $\{p_i,\rho_i\}$ is the optimal pure decomposition of $\rho$, $\rho_i=|\varphi_i\rangle\langle\varphi_i|$, then we see
\begin{equation*}
\begin{array}{rl}
C_{q-k}(\rho)&=\sum\limits_{i}p_{i}C_{q-k}(|\varphi_{i}\rangle)\\
&\geq\sum\limits_{i}p_{i}C_{q-k}(\rho_{i}^{\rm PI})\\
&\geq C_{q-k}(\rho^{\rm PI}).\\
\end{array}
\end{equation*}
Here the first inequality is based on inequality (\ref{21}), and the second inequality is due to the convexity of $C_{q-k}$.

Because the PI part depends on the choice of bases \cite{2} and the relations listed above are true for $\rho_{U}^{\rm PI}$ obtained under any locally unitary transformation $U$, one derives
\begin{equation*}
\begin{array}{rl}
C_{q-k}(\rho)\geq\max\limits_{U} C_{q-k}(\rho_{U}^{\rm PI}).
\end{array}
\end{equation*}

With the similar procedures, the inequality (\ref{13}) can be obtained.
$\hfill\blacksquare$

Here we only need to take into account the space of the permutationally invariant quantum states, instead of the whole space, which broadly reduces the dimension of the space to be considered. The structure of the multipartite quantum state is extremely complicated, so the approach introduced in Ref. \cite{2} provides great convenience for characterizing and detecting $k$-separability of general quantum states.

The following we will look for the relation between $q$-$k$-ME concurrence and global negativity \cite{38}. Global negativity, a measure between subsystem $p$ and the remaining subsystem, is given by
\begin{equation}\label{18}
\begin{array}{rl}
N^p=\frac{1}{d_p-1}(\|\rho^{T_{p}}\|_1-1)=-\frac{2}{d_p-1}\sum_i\lambda_i^{p-},
\end{array}
\end{equation}
where $\rho^{T_{p}}$ is partial transpose with respect to the subsystem $p$, $\|\cdot\|_1$ is trace norm, $\lambda_i^{p-}$ is negative eigenvalue of $\rho^{T_{p}}$, and $d_p$ denotes the dimension of subsystem $p$. When $d_p=2~(p=1,2,\cdots,n)$, the Eq. (\ref{18}) can be reduced to $N^p=\|\rho^{T_{p}}\|_1-1=-2\sum_i\lambda_i^{p-}$.

The lower bound of $q$-concurrence $(q\geq2)$ was derived by Yang $et~al$. using positive partial transpose (PPT) criterion and realignment criterion in Ref. \cite{5}. The relation is
\begin{equation}\label{19}
\begin{array}{rl}
C_q(\rho_{AB})\geq\frac{[\max\{\|\rho^{T_{A}}\|_1^{q-1},\|\mathcal{R}(\rho)\|_1^{q-1}\}-1]^2}{m^{2q-2}-m^{q-1}},
\end{array}
\end{equation}
where $\rho^{T_A}$ is partial transpose with regard to the subsystem $A$ and $\mathcal{R}$ is a realignment operation \cite{5}, and $m$ is obtained by taking the minimum of the dimensions of the two subsystems. On basis of the inequality (\ref{19}), we show the connection between $q$-$n$-ME concurrence $(q\geq2)$ and global negativity in the following theorem.

{\bf Theorem 2}. For any $n$-qubit quantum state $\rho$, the relation between the $q$-$n$-ME concurrence $(q\geq2)$ and global negativity of quantum state $\rho$ is as follows:
\begin{equation}\label{26}
\begin{array}{rl}
C_{q-n}(\rho)\geq\frac{\sum\limits_{k=1}^{n}[(N^k+1)^{q-1}-1]^2}{n(2^{2q-2}-2^{q-1})}.
\end{array}
\end{equation}
Here the inequality is saturated for $n$-qubit pure state when $q=2$.

{\bf Proof}. For any $n$-qubit pure state $\rho=|\varphi\rangle\langle\varphi|$,
\begin{equation*}
\begin{array}{rl}
C_{q-n}(\rho)=&\frac{\sum\limits_{k=1}^{n}[1-{\rm Tr}(\rho_{k}^q)]}{n}\\
=&\frac{C_{q 1|A^1}+C_{q 2|A^2}+\cdots+C_{q n|A^p}}{n}\\
\geq &\frac{[(N^1+1)^{q-1}-1]^2+\cdots+[(N^n+1)^{q-1}-1]^2}{n(2^{2q-2}-2^{q-1})}.\\
\end{array}
\end{equation*}
Here $A^p=\{1,2,\cdots,n\}\setminus\{p\}$, $p=1,2,\cdots,n$.

Note that when $q=2$, $C_{2p|A^p}(|\varphi\rangle)=\frac{[N(|\varphi\rangle)]^2}{2}$, then the inequality (\ref{26}) holds with equality when $q=2$.

For any $n$-qubit mixed state $\rho$, suppose that $\{p_i,\rho_i\}$ is the optimal pure decomposition and $\rho_i=|\phi_i\rangle\langle\phi_i|$, then
\begin{equation*}
\begin{array}{rl}
C_{q-n}(\rho)=&\sum\limits_{i}p_iC_{q-n}(|\phi_i\rangle)\\
\geq &\sum\limits_{i}p_i\frac{\sum\limits_{k=1}^{n}[(N^k(|\phi_i\rangle)+1)^{q-1}-1]^2}{n(2^{2q-2}-2^{q-1})}\\
\geq&\frac{\sum\limits_{k=1}^{n}[\sum\limits_{i}p_i(N^k(|\phi_i\rangle)+1)^{q-1}-1]^2}{n(2^{2q-2}-2^{q-1})}\\
\geq&\frac{\sum\limits_{k=1}^{n}[(\sum\limits_{i}p_i N^k(|\phi_i\rangle)+1)^{q-1}-1]^2}{n(2^{2q-2}-2^{q-1})}\\
\geq&\frac{\sum\limits_{k=1}^{n}[(N^k(\rho)+1)^{q-1}-1]^2}{n(2^{2q-2}-2^{q-1})},
\end{array}
\end{equation*}
where the second inequality holds because $y=x^2$ is a convex function, the third inequality is due to the fact that $y=x^{q-1}$ is convex for $q>2$, the third inequality is clearly true when $q=2$, and the last inequality holds from the convexity of $N^k$. $\hfill\blacksquare$

For higher dimensional systems, the result is as follows.

{\bf Corollary 1}. For any $n$-qudit quantum state $\rho\in\otimes_{i=1}^{n}\mathcal{H}_i$, ${\rm dim}\mathcal{H}_i=m$, $i=1,2,\cdots,n$, the relation between the $q$-$n$-ME concurrence $(q\geq2)$ and global negativity of quantum state $\rho$ is
\begin{equation}\label{029}
\begin{array}{rl}
C_{q-n}(\rho)\geq\frac{\sum\limits_{k=1}^{n}\{[(m-1)N^k+1]^{q-1}-1\}^2}{n(m^{2q-2}-m^{q-1})}.
\end{array}
\end{equation}

In Ref. \cite{37}, Wei $et~al$. provided lower bounds of $q$-concurrence for any bipartite state $\rho_{AB}$, which are
\begin{equation}\label{022}
\begin{array}{rl}
C_{q}(\rho_{AB})\geq\frac{1-m^{1-q}}{(m-1)^2}[\max(\|\rho^{T_A}\|_1, \|\mathcal{R}(\rho)\|_1)-1]^2\\
\end{array}
\end{equation}
for either $q\geq2$ with $m\geq3$ or $q\geq3$ with $m=2$, and
\begin{equation}\label{023}
\begin{array}{rl}
C_{q}(\rho_{AB})\geq\frac{1-2^{1-q}}{2-2^{2-s}}[\max(\|\rho^{T_A}\|_1, \|\mathcal{R}(\rho)\|_1)-1]^2\\
\end{array}
\end{equation}
for $2.4721=s\leq q<3$ with $m=2$. Here $m$ is the smallest of the dimensions of the two systems.

Next we improve the above results of Theorem 2 and Corollary 1 by utilizing the relation presented in formulas (\ref{022}) and (\ref{023}). For any $n$-qubit quantum state $\rho$, one derives
\begin{equation}\label{27}
\begin{array}{rl}
C_{q-n}(\rho)\geq\frac{2^{q-1}-1}{2^{q-1}n}{\sum_{k=1}^{n}(N^k)^2}
\end{array}
\end{equation}
for $q\geq3$, and
\begin{equation}
\begin{array}{rl}\label{28}
C_{q-n}(\rho)\geq\frac{1-2^{1-q}}{(2-2^{2-s})n}{\sum_{k=1}^{n}(N^k)^2}
\end{array}
\end{equation}
for $s\leq q<3$.

For any $n$-qudit quantum state $\rho\in\otimes_{i=1}^{n}\mathcal{H}_i$, ${\rm dim}\mathcal{H}_i=m$, $i=1,2,\cdots,n$, one gets
\begin{equation}\label{32}
\begin{array}{rl}
C_{q-n}(\rho)\geq\frac{1-m^{1-q}}{n(m-1)^2}{\sum_{k=1}^{n}(N^k)^2}
\end{array}
\end{equation}
for $q\geq2$.

For $0\leq\alpha\leq\frac{1}{2}$, we have the following result.

{\bf Theorem 3}. For any $n$-qubit quantum state $\rho$, the relation between the $\alpha$-$n$-ME concurrence $(0\leq\alpha\leq\frac{1}{2})$ and global negativity of quantum state $\rho$ is
\begin{equation}\label{29}
\begin{array}{rl}
C_{\alpha-n}(\rho)\geq\frac{2^{1-\alpha}-1}{n}{\sum_{k=1}^{n}N^k}.
\end{array}
\end{equation}
For any $n$-qudit quantum state $\rho\in\otimes_{i=1}^{n}\mathcal{H}_i$, ${\rm dim}\mathcal{H}_i=m$, $i=1,2,\cdots,n$, we obtain
\begin{equation}\label{34}
\begin{array}{rl}
C_{\alpha-n}(\rho)\geq\frac{m^{1-\alpha}-1}{n(m-1)}{\sum_{k=1}^{n}N^k}.\\
\end{array}
\end{equation}

Next we will use these bounds to detect the entangled states.

{\bf Example 1}. Consider the mixture of $n$-qubit W state and white noise
\begin{equation*}
\begin{array}{rl}
\rho=a{|W\rangle\langle W|}+\frac{1-a}{2^n}\mathbb{I},
\end{array}
\end{equation*}
where $|{W}\rangle=\frac{|0\cdots01\rangle+|0\cdots10\rangle+\cdots+|1\cdots00\rangle}{\sqrt n}$. By calculation, if $a\geq\frac{n}{n+2^n\sqrt{n-1}}$, then there is
\begin{equation*}
\begin{array}{rl}
N^1=N^2=\cdots=N^n=\frac{(2^n\sqrt{n-1}+n)a-n}{n2^{n-1}}.
\end{array}
\end{equation*}
Following from the relations presented in inequalities (\ref{27}), (\ref{28}), and (\ref{29}), for $a\in\Big[\frac{n}{n+2^n\sqrt{n-1}},1\Big]$, one can obtain
\begin{equation*}
\begin{array}{rl}
C_{q-n}(\rho)\geq\frac{2^{q-1}-1}{2^{q-1}}\Big(\frac{(2^n\sqrt{n-1}+n)a-n}{n2^{n-1}}\Big)^2
\end{array}
\end{equation*}
for $q\geq3$, and
\begin{equation*}
\begin{array}{rl}
C_{q-n}(\rho)\geq\frac{1-2^{1-q}}{2-2^{2-s}}\Big(\frac{(2^n\sqrt{n-1}+n)a-n}{n2^{n-1}}\Big)^2
\end{array}
\end{equation*}
for $s\leq q<3$, and
\begin{equation*}
\begin{array}{rl}
C_{\alpha-n}(\rho)\geq({2^{1-\alpha}-1})\frac{(2^n\sqrt{n-1}+n)a-n}{n2^{n-1}}
\end{array}
\end{equation*}
for $0\leq\alpha\leq\frac{1}{2}$.

When $a\in\Big(\frac{n}{n+2^n\sqrt{n-1}},1\Big]$, $\rho$ is $n$-nonseparable, that is, it is an entangled state. However, the result in Ref. \cite{19} is that the state $\rho$ is entangled if $a\in\Big(\frac{n}{2^n+n},1\Big]$. Due to $\frac{n}{n+2^n\sqrt{n-1}}<\frac{n}{2^n+n}$ when $n>2$, the range of entanglement that can be detected using the negativity method is larger than that in Ref. \cite{19}.

\section{The degree of separability}\label{V}
In Ref. \cite{17}, Hong $et~al$. gave two inequalities that can be used to determine whether a state is $k$-nonseparable. Let $|\phi_1\rangle=|0\rangle^{\otimes n}$ and $|\phi_2\rangle=|1\rangle^{\otimes n}$ for Theorem 3 in Ref. \cite{17}, if an $n$-qubit quantum state $\rho$ is $k$-separable, then it fulfills the inequality
\begin{equation}\label{40}
\begin{array}{rl}
(2^k-2)A\leq B,
\end{array}
\end{equation}
where
\begin{equation*}
\begin{array}{rl}
&A=|\rho_{1,2^n}|,\\
&B=\sum\limits_{i=2}^{2^n-1}\sqrt{\rho_{i,i}\rho_{2^n-i+1,2^n-i+1}}.\\
\end{array}
\end{equation*}
Let $|\psi_i^s\rangle=|0\rangle^{\otimes(i-1)}|1\rangle|0\rangle^{\otimes(n-i)}$  for Theorem 4 in Ref. \cite{17}, if an $n$-qubit quantum state $\rho$ is $k$-separable, then the result is accord with that of Ref. \cite{2}, which is
\begin{equation}\label{41}
\begin{array}{rl}
C\leq D+(n-k)E.
\end{array}
\end{equation}
Here
\begin{equation*}
\begin{array}{rl}
&C=\sum\limits_{0\leq i\neq j\leq n-1}|\rho_{2^i+1,2^j+1}|,\\
&D=\sum\limits_{0\leq i\neq j\leq n-1}\sqrt{\rho_{1,1}\rho_{2^i+2^j+1,2^i+2^j+1}},\\
&E=\sum\limits_{i=0}^{n-1}|\rho_{2^i+1,2^i+1}|.\\
\end{array}
\end{equation*}
Violation of any of the above inequalities (\ref{40}) and (\ref{41}) implies that the state is $k$-nonseparable.

Based on the relation of Eq. (\ref{40}), an effective $k_{\rm eff}^1$ can be defined,
\begin{equation}\label{42}
\begin{array}{rl}
k_{\rm eff}^1=\log_2(2+\frac{B}{A}).
\end{array}
\end{equation}
Note that although Eq. (\ref{42}) and Eq. (7) in Ref. \cite{43} have the same form of expression, they represent completely different meanings. $k_{\rm eff}^1$ quantifies the degree of separability, while Eq. (7) in Ref. \cite{43} reflects the degree of entanglement. Besides, the effective $k_{\rm eff}^2$ was presented in Ref. \cite{2}, which was obtained by inverting Eq. (\ref{41}),
\begin{equation}\label{43}
\begin{array}{rl}
k_{\rm eff}^2=n-\frac{C-D}{E}.
\end{array}
\end{equation}
So $k_{\rm eff}^2$ can be also regarded as the degree of separability. If an $n$-partite quantum state is $k$-separable, then we should have that $k_{\rm eff}^i\geq k$ $(i=1,2)$. We can say that if $k_{\rm eff}^1<k$ or $k_{\rm eff}^2<k$, then the quantum state is $k$-nonseparable.  In particular, for a fully separable quantum state, one ought to get that both $k_{\rm eff}^1$ and $k_{\rm eff}^2$ are greater than or equal to $n$. If one of the $k_{\rm eff}^i$ $(i=1,2)$ is less than 2, then the quantum state is genuinely entangled.

To illustrate our results more clearly, we present two concrete examples.

\begin{figure}[htbp]
\centering
{\includegraphics[width=9cm,height=7cm]{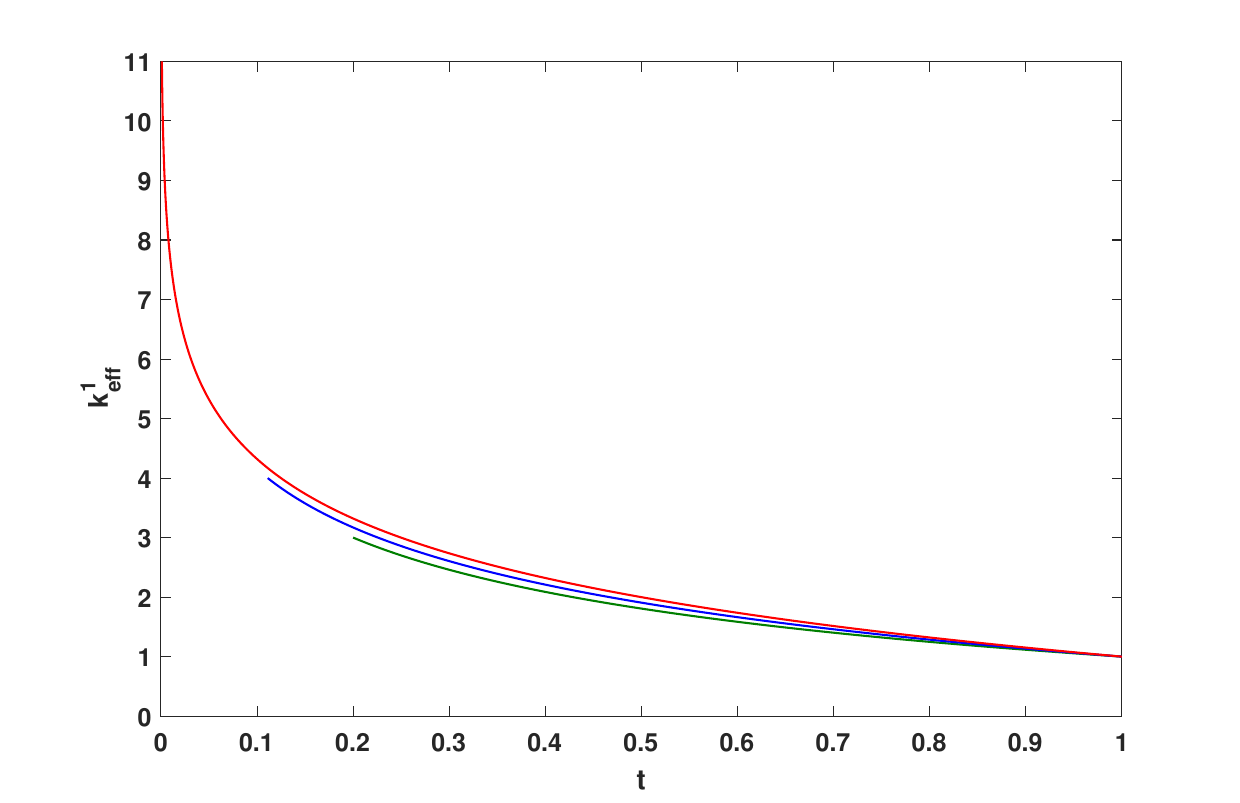}}
\caption{Interpretation the degree of separability $k_{\rm eff}^1(t,n)$ of the state given in formula (\ref{24}). The red line, blue line and green line, respectively, correspond to the cases where $n$ takes 11, 4, 3. Since $k_{\rm eff}^1>n$ is meaningless, it is omitted here.} \label{fig 5}
\end{figure}

{\bf Example 2}. Consider an $n$-qubit quantum state,
\begin{equation}\label{24}
\begin{array}{rl}
\rho=t{\rm|GHZ\rangle\langle GHZ|}+\frac{1-t}{2^n}\mathbb{I},\\
\end{array}
\end{equation}
where ${\rm|GHZ\rangle}=\frac{|0\rangle^{\otimes n}+|1\rangle^{\otimes n}}{\sqrt 2}$. By calculation, when $t\in\Big(\frac{2^n-2}{2^{n+k-1}-2},1\Big]$, the quantum state $\rho$ is $k$-nonseparable. That is, if $t\in\Big(\frac{2^n-2}{2^{n+k-1}-2},1\Big]$, then $C_{q-k}(\rho)>0$. Especially, if $t\in\big(\frac{2^n-2}{2^{n+1}-2},1\big]$, $\rho$ is not $2$-separable, namely, the state $\rho$ is genuinely multipartite entangled, which means $C_{q-2}(\rho)>0$. The state $\rho$ is full separable $(k=n)$ when $t\in[0,\frac{1}{2^{n-1}+1}]$, this range is consistent with the range given in Ref. \cite{39}. Due $\lim\limits_{n\rightarrow+\infty}\frac{2^n-2}{2^{n+1}-2}=\frac{1}{2}$, when $n$ enough large, nearly half of the states are genuinely multipartite entangled. Besides, $\lim\limits_{n\rightarrow+\infty}\frac{1}{2^{n-1}+1}=0$ which means the quantum states are almost entangled when $n\rightarrow+\infty$.  For this state, we can get
\begin{equation*}
\begin{array}{rl}
k_{\rm eff}^1(t,n)=\log_2\Big(\frac{(2^n-2)(1-t)}{2^{n-1}t}+2\Big)
\end{array}
\end{equation*}
which is a function of $t$ and $n$. Plotting $k_{\rm eff}^1(t,n)$, we take $n=3,4,11$ here. As shown in Fig. \ref{fig 5}, we observe that $k_{\rm eff}^1$ monotonically increasing with the amount of decoherence.

\begin{figure}[htbp]
\centering
{\includegraphics[width=9cm,height=7cm]{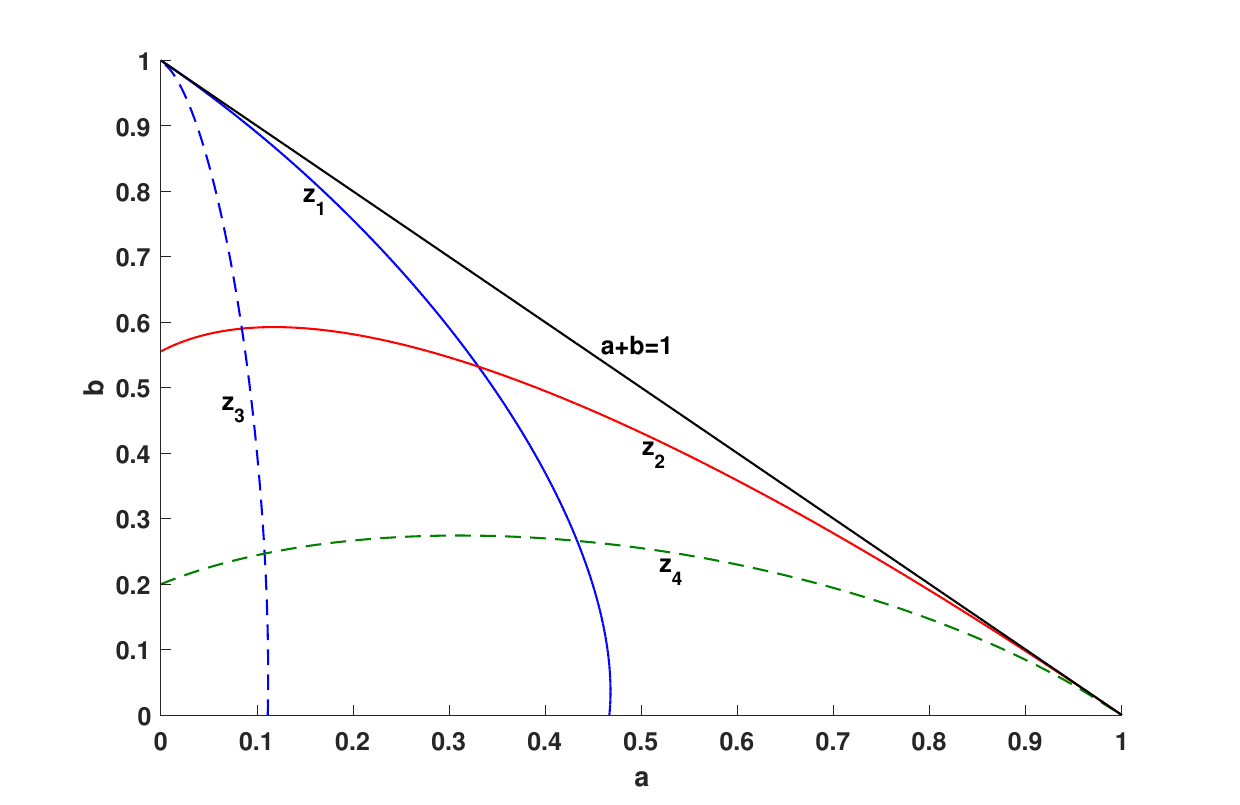}}
\caption{The solid blue line $z_1$ and the solid red line $z_2$ represent the first equality and the second equality for $n=4$ and $k=2$, respectively; the blue dashed line $z_3$ and the green dashed line $z_4$ denote the first equality and the second equality for $n=4$ and $k=4$, respectively. For case $a>b$, the inequality (\ref{40}) is better at detecting entanglement, whereas for case $a<b$, the inequality (\ref{41}) is better. Here the first equation and the second equation correspond to the formulas (\ref{40}) and (\ref{41}) taking equal signs, respectively.} \label{fig 1}
\end{figure}

{\bf Example 3}. Consider an $n$-qubit quantum state $\rho$, which is a mixture of GHZ state, W state, and white noise,
\begin{equation*}
\begin{array}{rl}
\rho=a{\rm|GHZ\rangle\langle GHZ|}+b{|W\rangle\langle W|}+\frac{1-a-b}{2^n}\mathbb{I},
\end{array}
\end{equation*}
where $|W\rangle=\frac{|0\cdots01\rangle+|0\cdots10\rangle+\cdots+|1\cdots00\rangle}{\sqrt n}$. By simple calculation, one has $A=\frac{a}{2}$, $B=2n\sqrt{\big(\frac{b}{n}+\frac{1-a-b}{2^n}\big)\frac{1-a-b}{2^n}}+(2^{n-1}-n-1)\frac{1-a-b}{2^{n-1}}$, $C=(n-1)b$, $D=n(n-1)\sqrt{\big(\frac{a}{2}+\frac{1-a-b}{2^n}\big)\frac{1-a-b}{2^n}}$, $E=n\big(\frac{b}{n}+\frac{1-a-b}{2^n}\big)$. Set $n=4$, when $k_{\rm eff}^1=\log_2[\frac{2}{a}\big(\frac{\sqrt{(1-a+3b)(1-a-b)}}{2}+\frac{3(1-a-b)}{8}\big)+2]<k$ or $k_{\rm eff}^2=4-\frac{12b-3\sqrt{(1+7a-b)(1-a-b)}}{1-a+3b}<k$, the quantum state $\rho$ is $k$-nonseparable. To make it more intuitive, we show the cases $k=2$ and $k=4$ in FIG. \ref{fig 1}.

As we can see in FIG. \ref{fig 1}, the states in the region bounded by line $z_1$, axis $a$, and line $a+b=1$, and the region bounded by line $z_2$, axis $b$, and line $a+b=1$ are genuinely 4-partite entangled. The states in the region bounded by line $z_1$,  $z_2$, and axis $a$ can only detected by the first form (\ref{40}), the states in the region bounded by line $z_1$,  $z_2$, and axis $b$ can only detected by the second form (\ref{41}), and the states in the intersection region bounded by line $z_1$, line $z_2$ and line $a+b=1$ are those where both inequalities can detect. Similarly, the states in the region bounded by line $z_3$, axis $a$, and line $a+b=1$, and the region bounded by line $z_4$, axis $b$, and line $a+b=1$ are not 4-separable, namely, they are entangled states.

Therefore, the combination of Eqs. (\ref{42}) and (\ref{43}) can be used to judge the separability of quantum states more effectively.

\section{Comparing $q$-GME concurrence with other GME measures}\label{VI}
We first introduce concurrence fill \cite{42} which is defined based on Heron formula of triangle area,
\begin{equation*}
\begin{array}{rl}
F_{A_1A_2A_3}=&\big[\frac{16}{3}P\big(P-C^2_{A_1|A_2A_3}\big)\big(P-C^2_{A_2|A_1A_3}\big)\\
&\big(P-C^2_{A_3|A_1A_2}\big)\big]^{\frac{1}{4}},\\
\end{array}
\end{equation*}
where $P=\frac{1}{2}\big(C^2_{A_1|A_2A_3}+C^2_{A_2|A_1A_3}+C^2_{A_3|A_1A_2}\big)$. $F_{A_1A_2A_3}$ perfectly characterizes the geometric meaning of three-qubit quantum states.

For any $n$-partite pure state $|\varphi\rangle$, the geometric mean of $q$-concurrence (G$q$C) \cite{40} is
\begin{equation*}
\begin{array}{rl}
\mathcal{G}_q(|\varphi\rangle)=[\mathcal{P}_q(|\varphi\rangle)]^{\frac{1}{c(\gamma)}},
\end{array}
\end{equation*}
where $\mathcal{P}_q(|\varphi\rangle)=\prod_{\gamma_i\in\gamma}C_{qA_{\gamma_i}|{\overline{A} _{\gamma_i}}}(|\varphi\rangle)$, $\gamma=\{\gamma_i\}$ denotes all of possible bipartitions, $c(\gamma)$ represents the cardinality of the set $\gamma$.

Mathematically, G$q$C is defined based on the geometric mean of bipartite concurrence, however, the $q$-GME concurrence is a special case of the $q$-$k$-ME concurrence defined in terms of the minimum of all possible $k$-partitions. Next we illustrate with a specific example exhibiting that $C_{q-\rm GME}$ and $\mathcal{G}_q$ are distinct. In fact, the $q$-GME concurrence is consistent with G$q$C for 3-partite completely symmetric pure states. For example, ${\rm | GHZ_3\rangle}=\frac{|000\rangle+|111\rangle}{\sqrt{2}}$ and ${|W_3\rangle}=\frac{|100\rangle+|010\rangle+|001\rangle}{\sqrt{3}}$,
$C_{q-\rm GME}({\rm |GHZ_3\rangle})=\mathcal{G}_q(\rm |GHZ_3\rangle)=1-\frac{1}{2^{q-1}}$ and $C_{q-\rm GME}({|W_3\rangle})=\mathcal{G}_q({ |W_3\rangle})=1-[(\frac{2}{3})^{q}+(\frac{1}{3})^{q}]$.

Now we compare these measures by a specific example. Theoretically, the $q$-GME concurrence defined by us may occur sharp peak due to the minimization involved, but here we will present an example to show that the measure defined by us is sometimes smoother than G$q$C and concurrence fill.

\begin{figure}[htbp]
\centering
{\includegraphics[width=9.5cm,height=6.5cm]{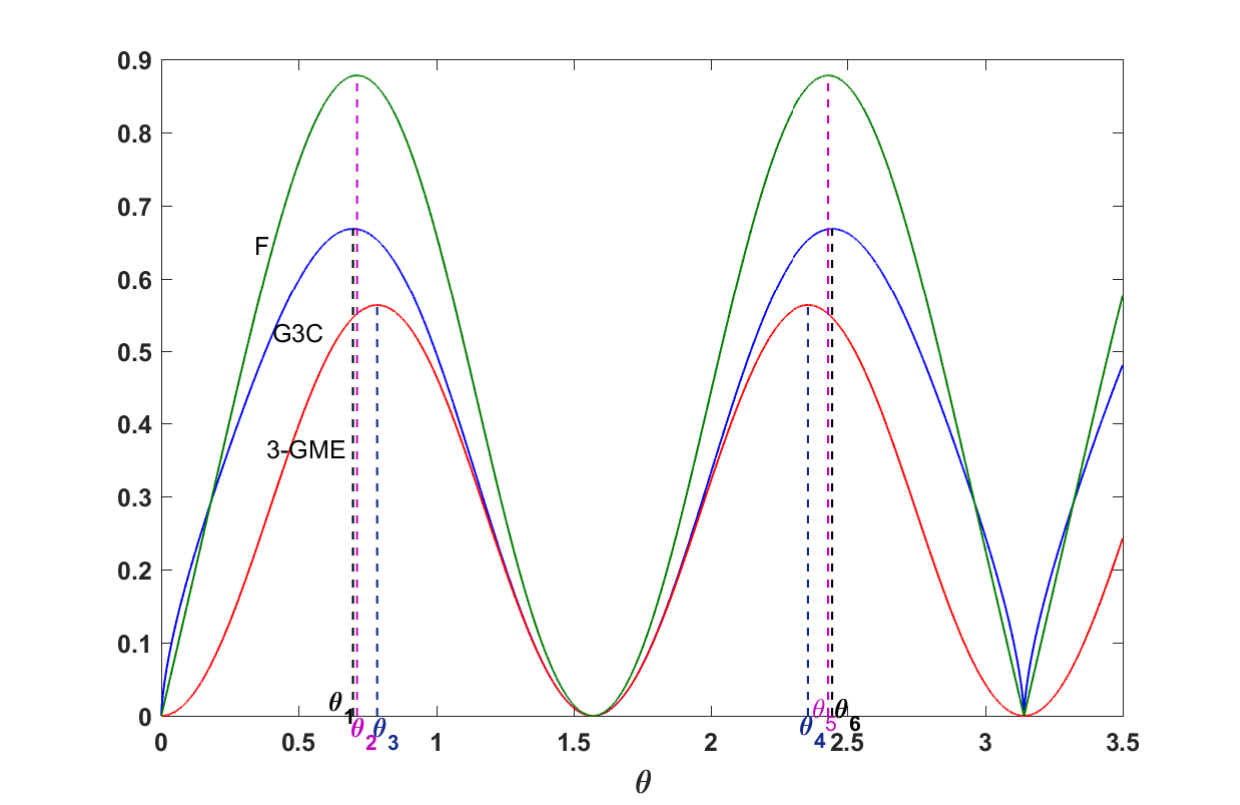}}
\caption{Set $q=3$. The blue line is $\mathcal{G}_3(|\phi\rangle)$; the red line expresses the $C_{3-{\rm GME}}(|\phi\rangle)$, the green line denotes concurrence fill of pure state $|\phi\rangle$.} \label{fig 3}
\end{figure}

{\bf Example 4}. Considering a quantum state $|\phi_\theta\rangle=-\frac{1}{2}{\rm cos}\theta|010\rangle+\frac{\sqrt{3}}{2}{\rm cos}\theta|100\rangle+{\rm sin}\theta|011\rangle$, we get
\begin{equation*}
\begin{array}{rl}
&C_{q-{\rm GME}}(|\phi_\theta\rangle)=\min\Big\{1-[(\frac{1}{4}+\frac{3}{4}{\rm sin}^2\theta)^q+(\frac{3}{4}{\rm cos}^2\theta)^q],1-\\
&~~~~~~~~~~~~~~\Big[\Big(\frac{1+\sqrt{1-3{\rm sin^2\theta}{\rm cos}^2\theta}}{2}\Big)^q+\Big(\frac{1-\sqrt{1-3{\rm sin^2\theta}{\rm cos}^2\theta}}{2}\Big)^q\Big]\Big\},\\
&\mathcal{G}_q(|\phi_\theta\rangle)=\Big\{1-[(\frac{1}{4}+\frac{3}{4}{\rm sin}^2\theta)^q+(\frac{3}{4}{\rm cos}^2\theta)^q]\Big\}^{\frac{2}{3}}\Big\{1-\\
&~~~~~~~~~~~~~~\Big[\Big(\frac{1+\sqrt{1-3{\rm sin^2\theta}{\rm cos}^2\theta}}{2}\Big)^q+\Big(\frac{1-\sqrt{1-3{\rm sin^2\theta}{\rm cos}^2\theta}}{2}\Big)^q\Big]\Big\}^{\frac{1}{3}},\\
&{F}(|\phi_\theta\rangle)=[\frac{16}{3}P(P-\mathcal{C}_1)^2(P-\mathcal{C}_2)]^{1/4},\\
\end{array}
\end{equation*}
where
\begin{equation*}
\begin{array}{rl}
&P=3-\Big\{\Big[\Big(\frac{1+\sqrt{1-3{\rm sin^2\theta}{\rm cos}^2\theta}}{2}\Big)^2+\Big(\frac{1-\sqrt{1-3{\rm sin^2\theta}{\rm cos}^2\theta}}{2}\Big)^2\Big]\\
&~~~~~~-2[(\frac{1}{4}+\frac{3}{4}{\rm sin}^2\theta)^2+(\frac{3}{4}{\rm cos}^2\theta)^2]\Big\},\\
&\mathcal{C}_1=2-2[(\frac{1}{4}+\frac{3}{4}{\rm sin}^2\theta)^2+(\frac{3}{4}{\rm cos}^2\theta)^2],\\
&\mathcal{C}_2=2-2\Big[\Big(\frac{1+\sqrt{1-3{\rm sin^2\theta}{\rm cos}^2\theta}}{2}\Big)^2+\Big(\frac{1-\sqrt{1-3{\rm sin^2\theta}{\rm cos}^2\theta}}{2}\Big)^2\Big].\\
\end{array}
\end{equation*}
When $q=3$, $C_{3-{\rm GME}}(|\phi_\theta\rangle)=1-\Big[\Big(\frac{1+\sqrt{1-3{\rm sin^2\theta}{\rm cos}^2\theta}}{2}\Big)^3+\Big(\frac{1-\sqrt{1-3{\rm sin^2\theta}{\rm cos}^2\theta}}{2}\Big)^3\Big]$. The comparison of three measures is shown in FIG. \ref{fig 3}.

Observing the FIG. \ref{fig 3}, we find the pure state $|\phi_\theta\rangle$ is not genuinely entangled when $\theta=0,\frac{\pi}{2},\pi$. When $\theta\in[\theta_2,\theta_3]\cup[\theta_4,\theta_5]$, the entanglement order of $C_{3-\rm GME}$ is different from G3C and concurrence fill, that is, there exist $\vartheta_1,~\vartheta_2\in[\theta_2,\theta_3]$ or $\vartheta_1,~\vartheta_2\in[\theta_4,\theta_5]$  such that $C_{3-{\rm GME}}(|\phi_{\vartheta_1}\rangle)\leq C_{3-{\rm GME}}(|\phi_{\vartheta_2}\rangle)$, while $\mathcal{G}_3(|\phi_{\vartheta_1}\rangle)\geq\mathcal{G}_3(|\phi_{\vartheta_2}\rangle)$ and ${F}(|\phi_{\vartheta_1}\rangle)\geq{F}(|\phi_{\vartheta_2}\rangle)$. In addition, when $\theta\in[\theta_1,\theta_2]\cup[\theta_5,\theta_6]$, the order of entanglement of $C_{3-\rm GME}$ and G3C is also different. When $\theta\in(0,\theta_3)$, only the measure we defined corresponds to unique quantum state. When $\theta=\pi$, the G3C and concurrence fill have a sharp peak, while our measure is smooth. Therefore, $q$-GME concurrence is advantageous in some cases.

\begin{figure}[htbp]
\centering
\subfigure[]{\includegraphics[width=9cm,height=6cm]{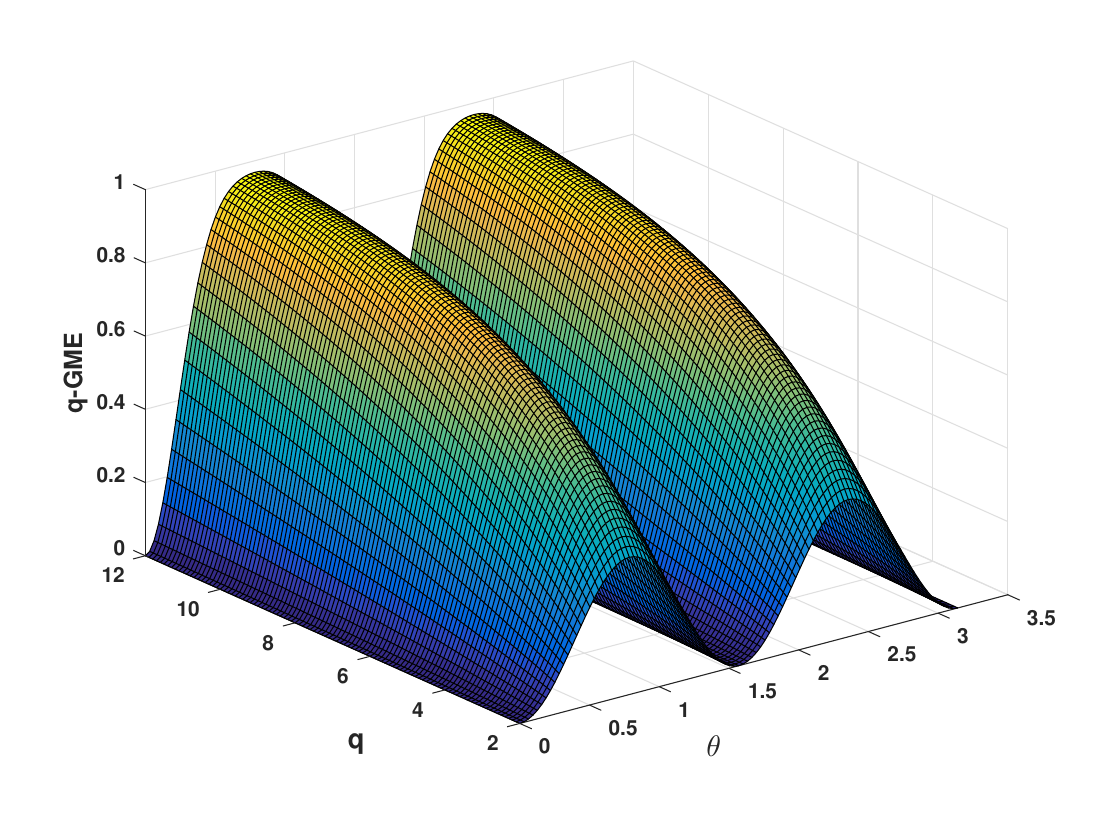}}
\quad
\subfigure[]{\includegraphics[width=9cm,height=6cm]{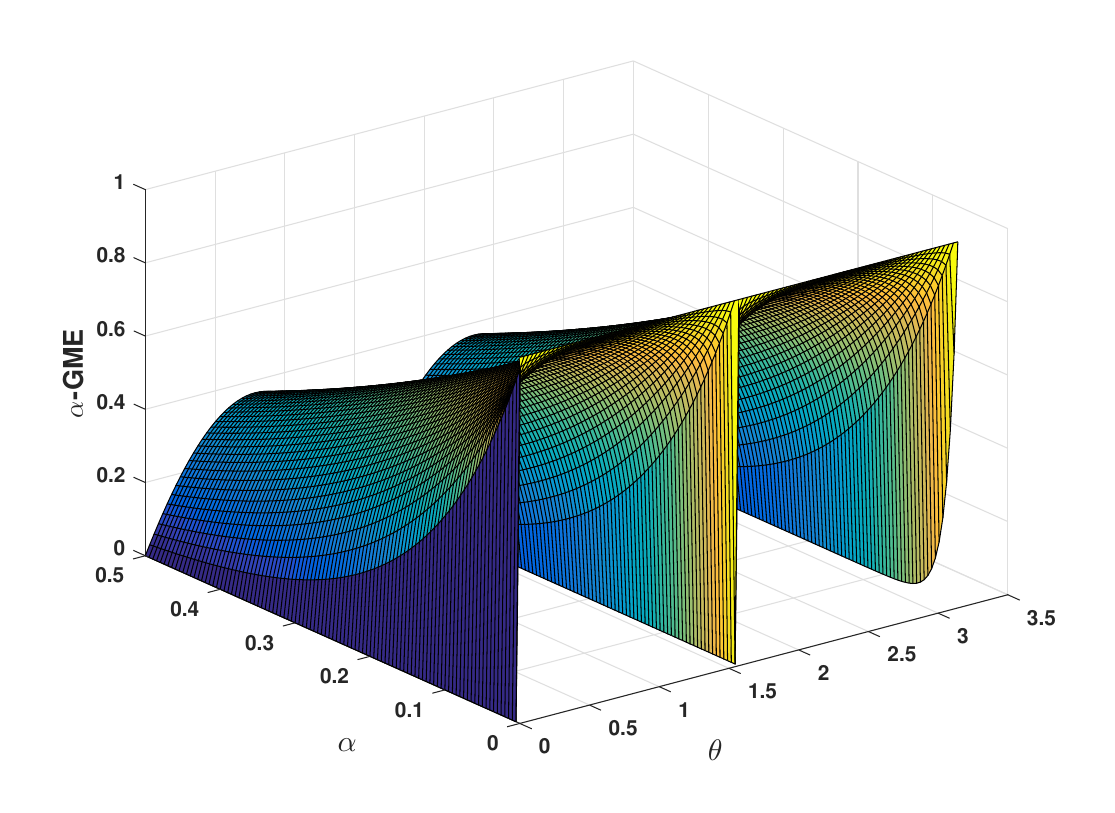}}
\caption{For $q\in[2,12]$, $C_{q-\rm GME}$ is an increasing function about $q$. For $\alpha\in[0,\frac{1}{2}]$, $C_{\alpha-\rm GME}$ is a decreasing function about $\alpha$.} \label{fig 4}
\end{figure}

To see whether $C_{q-{\rm GME}}(|\phi\rangle)$ and $C_{\alpha-{\rm GME}}(|\phi\rangle)$ have the same monotonicity about $q$ and $\alpha$, respectively, we plot two figures. In (a) of FIG. \ref{fig 4}, we take $q\in[2,12]$, the function $C_{q-\rm GME}$ is increasing of $q$. In (b) of FIG. \ref{fig 4}, we take $\alpha\in[0,\frac{1}{2}]$, the function $C_{\alpha-{\rm GME}}$ is decreasing of $\alpha$. This also reflects that $C_{q-k}$ and $C_{\alpha-k}$ are different.

\section{conclusion}\label{VII}
In this work, we have proposed two types of general parameterized entanglement measures, $q$-$k$-ME concurrence $C_{q-k}$ $(q\geq2,~2\leq k\leq n)$ and $\alpha$-$k$-ME concurrence $C_{\alpha-k}$ $(0\leq\alpha\leq\frac{1}{2},~2\leq k\leq n)$, in $n$-partite systems from the standpoint of $k$-nonseparability, and shown that these measures $C_{q-k}$ and $C_{\alpha-k}$  satisfy the requirements (M1)---(M5). Besides, $C_{q-k}$ also satisfy the property (M6). It is evident that, as special cases of $C_{q-k}$ and $C_{\alpha-k}$, parameterized GME measures $C_{q-\rm GME}$ and $C_{\alpha-\rm GME}$ can inherit their properties, respectively. The $q$-$k$-ME concurrence of $\rho$ is lower bounded by the maximum of $C_{q-k}$ of the PI part of $\rho$, and so is $C_{\alpha-k}(\rho)$. Apart from that, we have associated global negativity with $q$-$n$-ME concurrence ($\alpha$-$n$-ME concurrence), and gave the lower bound of $C_{q-n}$ ($C_{\alpha-n}$) which could be used to detect whether a quantum state is entangled. We have presented an example which is a mixture of W state and white noise, and observed that when $a\in\Big(\frac{n}{n+2^n\sqrt{n-1}},\frac{n}{2^n+n}\Big]$, our result can detect that these states are entangled, whereas the result in Ref. \cite{19} cannot. What's more, we have discussed the degree of entanglement of  $k$-nonseparable states, where the violation of any of these inequalities (\ref{40}) and (\ref{41}) implies that the state is $k$-nonseparable. The combination of these two formulas can detect entanglement more effectively. Comparing the $q$-GME concurrence with G$q$C and concurrence fill through a concrete example, we found that they generate different entanglement orders and that $q$-GME concurrence is sometimes smooth. The measures defined by us could be useful for further study of multipartite quantum entanglement.

\section*{ACKNOWLEDGMENTS}
This work was supported by the National Natural Science Foundation of China under Grants No. 12071110 and No. 62271189, the Science and Technology Project of Hebei Education Department under Grant No. ZD2021066, and funded by School of Mathematical Sciences of Hebei Normal University under Grant No. xycxzz2023002.


\begin{thebibliography}{99}
\bibitem{9} A. K. Ekert, Quantum cryptography based on Bell's theorem,
    \href{https://journals.aps.org/prl/abstract/10.1103/PhysRevLett.67.661} {Phys. Rev. Lett. \textbf{67}, 661 (1991)}.
\bibitem{10} C. H. Bennett, G. Brassard, and N. D. Mermin, Quantum cryptography without Bell's theorem,
    \href{https://journals.aps.org/prl/abstract/10.1103/PhysRevLett.68.557} {Phys. Rev. Lett. \textbf{68}, 557 (1992)}.
\bibitem{8} B. A. Slutsky, R. Rao, P. C. Sun, and Y. Fainman, Security of quantum cryptography against individual attacks,
    \href{https://journals.aps.org/pra/abstract/10.1103/PhysRevA.57.2383} {Phys. Rev. A \textbf{57}, 2383 (1998)}.
\bibitem{7} L. Masanes, Universally composable privacy amplification from causality constraints,
    \href{https://journals.aps.org/prl/abstract/10.1103/PhysRevLett.102.140501} {Phys. Rev. Lett. \textbf{102}, 140501 (2009)}.
\bibitem{11} C. H. Bennett, G. Brassard, C. Cr\'{e}peau, R. Jozsa, A. Peres, and W. K. Wootters, Teleporting an unknown quantum state via dual classical and Einstein-Podolsky-Rosen channels,
    \href{https://journals.aps.org/prl/abstract/10.1103/PhysRevLett.70.1895} {Phys. Rev. Lett. \textbf{70}, 1895 (1993)}.
\bibitem{12} F. L. Yan and X. Q. Zhang, A scheme for secure direct communication using EPR pairs and teleportation,
    \href{https://link.springer.com/article/10.1140/epjb/e2004-00296-4} {Eur. Phys. J. B \textbf{41}, 75 (2004)}.
\bibitem{13} T. Gao, F. L. Yan, and Y. C. Li, Optimal controlled teleportation,
    \href{https://iopscience.iop.org/article/10.1209/0295-5075/84/50001} {Europhys. Lett. \textbf{84}, 50001 (2008)}.
\bibitem{14}  J. M. Renes and M. Grassl, Generalized decoding, effective channels, and simplified security proofs in quantum key distribution,
    \href{https://journals.aps.org/pra/abstract/10.1103/PhysRevA.74.022317} {Phys. Rev. A \textbf{74}, 022317 (2006)}.
\bibitem{15} Y. H. Zhou, Z. W. Yu, and X. B. Wang, Making the decoy-state measurement-device-independent quantum key distribution practically useful,
    \href{https://journals.aps.org/pra/abstract/10.1103/PhysRevA.93.042324} {Phys. Rev. A \textbf{93}, 042324 (2016)}.
\bibitem{6} F. L. Yan, T. Gao, and E. Chitambar, Two local observables are sufficient to characterize maximally entangled states of $N$ qubits,
    \href{https://journals.aps.org/pra/abstract/10.1103/PhysRevA.83.022319} {Phys. Rev. A \textbf{83}, 022319 (2011)}.


\bibitem{23} S. A. Hill and W. K. Wootters, Entanglement of a pair of quantum bits,
    \href{https://journals.aps.org/prl/abstract/10.1103/PhysRevLett.78.5022} {Phys. Rev. Lett. \textbf{78}, 5022 (1997)}.
\bibitem{24} W. K. Wootters, Entanglement of formation of an arbitrary state of two qubits,
    \href{https://journals.aps.org/prl/abstract/10.1103/PhysRevLett.80.2245} {Phys. Rev. Lett. \textbf{80}, 2245 (1998)}.
\bibitem{25} F. Mintert, M. Ku\'{s}, and A. Buchleitner, Concurrence of mixed multipartite quantum states,
    \href{https://journals.aps.org/prl/abstract/10.1103/PhysRevLett.95.260502} {Phys. Rev. Lett. \textbf{95}, 260502 (2005)}.
\bibitem{22} R. Horodecki, P. Horodecki, M. Horodecki, and K. Horodecki, Quantum entanglement,
    \href{https://journals.aps.org/rmp/abstract/10.1103/RevModPhys.81.865} {Rev. Mod. Phys. \textbf{81}, 865 (2009)}.
\bibitem{26} K. \.{Z}yczkowski, P. Horodecki, A. Sanpera, and M. Lewenstein, Volume of the set of separable states,
    \href{https://journals.aps.org/pra/abstract/10.1103/PhysRevA.58.883} {Phys. Rev. A \textbf{58}, 883 (1998)}.
\bibitem{27} G. Vidal and R. F. Werner, Computable measure of entanglement,
    \href{https://journals.aps.org/pra/abstract/10.1103/PhysRevA.65.032314} {Phys. Rev. A \textbf{65}, 032314 (2002)}.
\bibitem{28} C. H. Bennett, D. P. DiVincenzo, J. A. Smolin, and W. K. Wootters, Mixed-state entanglement and quantum error correction,
    \href{https://journals.aps.org/pra/abstract/10.1103/PhysRevA.54.3824} {Phys. Rev. A \textbf{54}, 3824 (1996)}.
\bibitem{29} P. W. Shor, Equivalence of additivity questions in quantum information theory,
    \href{https://link.springer.com/article/10.1007/s00220-003-0981-7} {Commun. Math. Phys. \textbf{246}, 453 (2004)}.
\bibitem{30} J. S. Kim, Tsallis entropy and entanglement constraints in multiqubit systems,
    \href{https://journals.aps.org/pra/abstract/10.1103/PhysRevA.81.062328} {Phys. Rev. A \textbf{81}, 062328 (2010)}.
\bibitem{31} P. Krammer, H. Kampermann, D. Bru{\ss}, R. A. Bertlmann, L. C. Kwek, and C. Macchiavello, Multipartite entanglement detection via structure factors,
    \href{https://journals.aps.org/prl/abstract/10.1103/PhysRevLett.103.100502} {Phys. Rev. Lett. \textbf{103}, 100502 (2009)}.


\bibitem{32} O. G\"{u}hne and M. Seevinck, Separability criteria for genuine multiparticle entanglement,
    \href{https://iopscience.iop.org/article/10.1088/1367-2630/12/5/053002} {New J. Phys. \textbf{12}, 053002 (2010)}.
\bibitem{34} O. Gittsovich, P. Hyllus, and O. G\"{u}hne, Multiparticle covariance matrices and the impossibility of detecting graph-state entanglement with two-particle correlations,
    \href{https://journals.aps.org/pra/abstract/10.1103/PhysRevA.82.032306} {Phys. Rev. A \textbf{82}, 032306 (2010)}.
\bibitem{3} T. Gao and Y. Hong, Detection of genuinely entangled and nonseparable $n$-partite quantum states,
    \href{https://journals.aps.org/pra/abstract/10.1103/PhysRevA.82.062113} {Phys. Rev. A \textbf{82}, 062113 (2010)}.
\bibitem{33} M. Huber, H. Schimpf, A. Gabriel, C. Spengler, D. Bru{\ss}, and B. C. Hiesmayr, Experimentally implementable criteria revealing substructures of genuine multipartite entanglement,
    \href{https://journals.aps.org/pra/abstract/10.1103/PhysRevA.83.022328} {Phys. Rev. A \textbf{83}, 022328 (2011)}.
\bibitem{18} T. Gao and Y. Hong, Separability criteria for several classes of $n$-partite quantum states,
     \href{https://link.springer.com/article/10.1140/epjd/e2010-10432-4} {Eur. Phys. J. D \textbf{61}, 765 (2011)}.
\bibitem{19} T. Gao, Y. Hong, Y. Lu, and F. L. Yan,  Efficient $k$-separability criteria for mixed multipartite quantum states,
    \href{https://iopscience.iop.org/article/10.1209/0295-5075/104/20007} {Europhys. Lett. \textbf{104}, 20007 (2013)}.
\bibitem{17} Y. Hong, T. Gao, and F. L. Yan, Detection of $k$-partite entanglement and $k$-nonseparability of multipartite quantum states,
    \href{https://www.sciencedirect.com/science/article/pii/S0375960121002115?via} {Phys. Lett. A \textbf{401}, 127347 (2021)}.
\bibitem{36} Y. Hong, X. F. Qi, T. Gao, and F. L. Yan, Detection of multipartite entanglement via quantum Fisher information,
    \href{https://iopscience.iop.org/article/10.1209/0295-5075/134/60006} {Europhys. Lett. \textbf{134}, 60006 (2021)}.
\bibitem{35} Y. Hong, X. F. Qi, T. Gao, and F. L. Yan, Detection of the quantum states containing at most $k-1$  unentangled particles,
    \href{https://iopscience.iop.org/article/10.1088/1674-1056/abfb5e} {Chin. Phys. B \textbf{30}, 100306 (2021)}.
\bibitem{39} L. M. Zhang, T. Gao, and F. L. Yan, Relations among $k$-ME concurrence, negativity, polynomial invariants, and tangle,
    \href{https://link.springer.com/article/10.1007/s11128-019-2223-8} {Quantum Inf. Process. \textbf{18}, 194 (2019)}.


\bibitem{20} Z. H. Ma, Z. H. Chen, J. L. Chen, C. Spengler, A. Gabriel, and M. Huber, Measure of genuine multipartite entanglement with computable lower bounds,
    \href{https://journals.aps.org/pra/abstract/10.1103/PhysRevA.83.062325} {Phys. Rev. A \textbf{83}, 062325 (2011)}.
\bibitem{21} Z. H. Chen, Z. H. Ma, J. L. Chen, and S. Severini, Improved lower bounds on genuine-multipartite-entanglement concurrence,
    \href{https://journals.aps.org/pra/abstract/10.1103/PhysRevA.85.062320} {Phys. Rev. A \textbf{85}, 062320 (2012)}.
\bibitem{1}  Y. Hong, T. Gao, and F. L. Yan, Measure of multipartite entanglement with computable lower bounds,
    \href{https://journals.aps.org/pra/abstract/10.1103/PhysRevA.86.062323} {Phys. Rev. A \textbf{86}, 062323 (2012)}.
\bibitem{2}  T. Gao, F. L. Yan, and S. J. van Enk, Permutationally invariant part of a density matrix and nonseparability of
$N$-qubit states,
    \href{https://journals.aps.org/prl/abstract/10.1103/PhysRevLett.112.180501}  {Phys. Rev. Lett. \textbf{112}, 180501 (2014)}.
\bibitem{42} S. Xie and J. H. Eberly, Triangle measure of tripartite entanglement,
    \href{https://journals.aps.org/prl/abstract/10.1103/PhysRevLett.127.040403} {Phys. Rev. Lett. \textbf{127}, 040403 (2021)}.
\bibitem{41} Y. F. Li and J. W. Shang, Geometric mean of bipartite concurrences as a genuine multipartite entanglement measure,
    \href{https://journals.aps.org/prresearch/abstract/10.1103/PhysRevResearch.4.023059} {Phys. Rev. Research \textbf{4}, 023059 (2022)}.
\bibitem{5} X. Yang, M. X. Luo, Y. H. Yang, and S. M. Fei, Parametrized entanglement monotone,
    \href{https://journals.aps.org/pra/abstract/10.1103/PhysRevA.103.052423} {Phys. Rev. A \textbf{103}, 052423 (2021)}.
\bibitem{4} G. Vidal, Entanglement monotones,
    \href{https://www.tandfonline.com/doi/abs/10.1080/09500340008244048} {J. Mod. Opt. \textbf{47}, 355 (2000)}.
\bibitem{16} Z. W. Wei and S. M. Fei, Parameterized bipartite entanglement measure,
    \href{https://iopscience.iop.org/article/10.1088/1751-8121/ac7592} {J. Phys. A: Math. Theor. \textbf{55}, 275303 (2022)}.
\bibitem{40} X. Shi, A genuine multipartite entanglement measure generated by the parametrized entanglement measure,
    \href{https://arxiv.org/pdf/2206.02232.pdf} {arXiv:2206.02232}.

\bibitem{37} Z. W. Wei, M. X. Luo, and S. M. Fei, Estimating parameterized entanglement measure,
    \href{https://link.springer.com/article/10.1007/s11128-022-03551-4} {Quantum Inf. Process. \textbf{21}, 210 (2022)}.
\bibitem{38} S. S. Sharma and N. K. Sharma, Quantum coherences, $K$-way negativities and multipartite entanglement,
    \href{https://journals.aps.org/pra/abstract/10.1103/PhysRevA.77.042117} {Phys. Rev. A \textbf{77}, 042117 (2008)}.
\bibitem{43} Y. Hong, X. F. Qi, T. Gao, and F. L. Yan, A $(k+1)$-partite entanglement measure of $N$-partite quantum states,
    \href{https://arxiv.org/abs/2211.03266} {arXiv:2211.03266}.



\end{thebibliography}

\end{document}